\documentclass{article}
\usepackage[dvips]{graphicx}
\usepackage{fullpage}
\newcommand\beq{\begin{equation}}
\newcommand\eeq{\end{equation}}
\newcommand\bea{\begin{eqnarray}}
\newcommand\eea{\end{eqnarray}}
\newcommand\beano{\begin{eqnarray*}}
\newcommand\eeano{\end{eqnarray*}}
\newcommand\LL{{\cal L}}
\newcommand\eps{\epsilon}
\newcommand\al{\alpha}
\newcommand\be{{\beta}}
\newcommand\ga{{\gamma}}
\newcommand\om{{\omega}}
\newcommand\ka{{\kappa}}
\newcommand\eqref[1]{(\ref{#1})}

\begin{document}


\newcommand{\qb}{$Q$-ball}

\begin{flushright}
\vskip-0.75cm
UdeM-GPP-TH-06-146\\
\end{flushright}
\bigskip

\begin{center}
{\LARGE $Q$-balls in Maxwell-Chern-Simons theory}\\
\bigskip\bigskip
M.~Deshaies-Jacques and R.~MacKenzie\\
\it Physique des particules, Universit\'e de Montr\'eal\\
C.P. 6128, Succ. Centre-ville, Montr\'eal, QC H3C 3J7
\end{center}

\bigskip
\begin{abstract}
We examine the energetics of $Q$-balls in Maxwell-Chern-Simons theory in
two space dimensions. Whereas gauged $Q$-balls are unallowed in this dimension in the absence of a Chern-Simons term due to a divergent electromagnetic energy, the addition of a Chern-Simons term introduces a gauge field mass and renders finite the otherwise-divergent electromagnetic energy of the $Q$-ball. Similar to the case of gauged $Q$-balls, Maxwell-Chern-Simons $Q$-balls have a maximal charge. The properties of these solitons are studied as a function of the parameters of the model considered, using a numerical technique known as relaxation. The results are compared to expectations based on qualitative arguments.
\end{abstract}
\bigskip


A class of non-topological solitons (see \cite{Lee:1991ax} for a comprehensive review) dubbed {\qb}s were examined some time ago by Coleman \cite{Coleman:1985ki}. These objects owe their existence to a conserved global charge. Under certain circumstances, a localized configuration of charge $Q$ can be created which has a lower energy than the ``naive" lowest-energy configuration, namely, $Q$ widely-separated ordinary particles (each of unit charge) at zero momentum. This latter state obviously has energy $Qm$, where $m$ is the mass of the quanta of the theory.
If another configuration of charge $Q$ can be constructed whose energy is lower, then that state cannot decay into ordinary matter: either it is stable or some other ``non-naive" configuration of the same charge and still lower energy is stable.

Coleman studied the three-dimensional case with a charged scalar field $\phi$. The {\qb} is a spherically symmetric configuration where $|\phi|$ is nonzero inside a core region and tends to zero as $r\to\infty$; its phase varies linearly in time. A mechanical analogy permits a clean demonstration of the necessary conditions which must be satisfied by the potential in order for {\qb}s to exist. It is particularly easy to analyze the case of large charge, since then the surface energy can be neglected compared to the volume energy. Among Coleman's conclusions is the fact that there is no upper limit to the charge of a {\qb} (if they exist in the first place); furthermore, the interior of a sufficiently large {\qb} is homogeneous.

The case of small {\qb}s was analyzed by Kusenko \cite{Kusenko:1997ad}, who (along with many others) also proposed possible astrophysical signatures of {\qb}s (see \cite{Kusenko:1999gz} and references therein). Possible applications in condensed matter physics have been analyzed in \cite{Hong:1987ur,Enqvist:2003zb,Nussinov:2004hf}. A number of other variations have been studied since, including non-abelian {\qb}s \cite{Safian:1987pr,Safian:1988cz}, gauged {\qb}s \cite{Lee:1988ag,Anagnostopoulos:2001dh}, $Q$-stars \cite{Lynn:1988rb}, {\qb}s in other dimensions \cite{Prikas:2004fx,Gleiser:2005iq}, higher-dimensional $Q$-objects \cite{MacKenzie:2001av,Axenides:2001pi}, spinning {\qb}s \cite{Volkov:2002aj,Kleihaus:2005me}, and so on.

Of particular interest here is the paper of Lee, et al. \cite{Lee:1988ag}, who discussed the case of gauged {\qb}s in three dimensions, using a combination of analytical and numerical techniques. They argued that when the charge exceeds a critical value the {\qb}'s energy exceeds $Qm$, so the \qb\ is at best metastable. This is intuitively reasonable, since a ball of electric charge will have a Coulomb energy which grows roughly as the square of the charge, so eventually the \qb\ will be unable to compete with ordinary matter. On the other hand, as the charge decreases the {\qb} gets smaller and smaller; surface effects become important and eventually destabilize the {\qb}. These two observations indicate that there may or may not be a range of charges for which {\qb}s exist, depending on under what circumstances each effect becomes significant.

Another result of their analysis is that the core of a large gauged {\qb} is not homogeneous, essentially because the charge repels itself, and the electromagnetic energy is reduced by having the charge migrate to the surface of the \qb.

In two dimensions, the gauged \qb's existence is problematic, for a fairly straightforward reason: the electric field goes like $1/r$ and the electric field energy diverges logarithmically. Such a divergence is sufficiently mild that one could still contemplate a configuration of several positively- and negatively-charged {\qb}s with total charge neutrality, in the spirit of global cosmic strings and vortices in liquid helium which also have logarithmically divergent energies. Nonetheless, strictly speaking, an isolated gauged {\qb} in two dimensions has divergent energy and therefore cannot hope to compete energetically with ordinary matter.

However, a new possibility exists in two dimensions: one can consider a model with a Chern-Simons term, either on its own or in addition to the usual Maxwell term.  The motivation for studying such a model falls into two classes. First, at least two concrete physical examples where a Chern-Simons term has important effects on a planar system have been advanced: the fractional quantum Hall effect (see \cite{Prange:1990} for a thorough discussion), and a proposed mechanism of superconductivity based on anyons \cite{Wilczek:1990ik}. In addition, whenever a fairly simple model gives rise to such rich behaviour, it is worth examining in detail, without the need to evoke physical applications to justify the work.

 Among the well-known physical effects of the Chern-Simons term (to say nothing of its profound mathematical properties) are a greater interplay between electric and magnetic phenomena (for example, a static charge distribution gives rise to both electric and magnetic fields) \cite{MacKenzie:1988ft,Goldhaber:1988iw}, parity and time reversal violation \cite{Redlich:1983kn}, fractional spin and statistics \cite{Wilczek:1983cy,Arovas:1985yb}, and mass generation for the gauge field \cite{Deser:1981wh}. The latter property is particularly pertinent here since the electric field of a Maxwell-Chern-Simons (MCS) {\qb} decays exponentially, and the argument given above leading to the conclusion that the electric field energy diverges no longer applies. Thus, we can address the question of whether gauged {\qb}s exist (that is to say, whether they can compete energetically with ordinary matter) in 2 space dimensions if the Chern-Simons term is present.

The possibility of MCS {\qb}s was noted by Khare and Rao \cite{Khare:1989dx,Khare:1990jz}, who argued that finite-energy charged configurations can exist in such models. However, they did not study the energetics to see how these configurations compare in energy to ordinary matter. Nontopological solitons similar to {\qb}s have been studied in certain self-dual models with Chern-Simons term in \cite{Jackiw:1990pr,Lee:1990eq}; see also \cite{Jackiw:1991au} for a review.

In this paper we do such an analysis, numerically. We begin by describing the model studied and the \qb\ ansatz generalized to the MCS case. Next, we make some qualitative observations to indicate what we might expect. In the remainder of the paper, we describe the numerical approach used and the results obtained.

The model we consider has a complex scalar field $\phi(x)$ with gauged U(1) symmetry, described by the following Lagrangian (in 2+1 dimensions):
\beq
\LL=-\frac14 F_{\mu\nu}^2+{\kappa\over2}\eps^{\al\be\ga}A_\al\partial_\be A_\ga
+|D_\mu\phi|^2
-V(\phi),
\label{one}
\eeq
where $F_{\mu\nu}=\partial_\mu A_\nu - \partial_\nu A_\mu$, $D_\mu\phi=(\partial_\mu+ie A_\mu)\phi$, and the potential is
\beq
V(\phi)=\phi^*\phi-\frac12(\phi^*\phi)^2+\frac{g}3(\phi^*\phi)^3.
\label{two}
\eeq
We have eliminated the coefficients of the quadratic and quartic terms in the potential with appropriate rescalings of the coordinate and fields. The potential is renormalizable in 2+1 dimensions, and is the simplest which admits ungauged {\qb}s \cite{Coleman:1985ki}. The precise requirements of the potential are that it be minimized at the origin (so that the symmetry is unbroken), and that a parabola passing through the origin exists which, firstly, is wider than the potential at the origin, and, secondly, intersects the potential at some nonzero field value. These are satisfied if $g>3/16$.

The conserved particle number (henceforth referred to as charge) is
\beq
Q=i\int d^2x \,\phi^*{\stackrel\leftrightarrow{D}}_0\phi,
\label{three}
\eeq
while the energy is
\beq
E=\int d^2x\left( |D_0\phi|^2 + |D_i\phi|^2+V(\phi)+\frac12F_{0i}^2+\frac14F_{ij}^2  \right).
\label{four}
\eeq

Since the mass of the $\phi$ field  is unity, this sets the standard to which {\qb}s must be compared. If we can construct a field configuration for which $E/Q<1$, then it cannot decay into ordinary matter, and either it or some other lower-energy configuration of the same charge is stable. To look for such a configuration, we use an ansatz where the scalar field is rotationally symmetric and has a constant frequency $\om$, along with appropriate gauge fields (recall that, with the Chern-Simons term, electric charge is a source for both electric and magnetic fields):
\beq
\phi(x)=e^{-i\om t}f(r),\qquad
A^0(x)=\al(r),\qquad
A^i(x)= {\eps^{ij}r_j\over r}\beta(r).
\label{five}
\eeq
The field equations become
\bea
f'' + {f'\over r}+\left((\om-e\al)^2-e^2\beta^2 -1\right)f +f^3-gf^5&=&0,\nonumber\\
\al''+{\al'\over r}-\kappa(\be'+{\be\over r})+2e(\om-e\al)f^2&=&0,
\label{six}\\
\be''+{\be'\over r}-{\be\over r^2}-\kappa \al' -2e^2\be f^2&=&0,\nonumber
\eea
while the charge and energy are

\parbox{10cm}{
\beano
Q&=&\int d^2x\,2(\om-e\al)f^2,   \\
E&=&\int d^2x\,\left\{ (\om-e\al)^2f^2+f'^2+e^2\be^2f^2+f^2-{1\over2}f^4+{g\over3}f^6
+{\al'^2+(\be'+\be/r)^2\over2}\right\}.\\
\eeano}
\hfill
\parbox{1cm}{
\beq
\label{seven}
\eeq}

\noindent The boundary conditions at the origin are
\beq
f'(0)=0,\qquad \al'(0)=0,\qquad \be(0)=0,
\label{eight}
\eeq
while as $r\to\infty$ these three fields must tend towards zero. Their asymptotic behavior in this limit is
\beq
f(r)\sim e^{-\sqrt{1-\om^2}r},\qquad \al(r)\sim e^{-\ka r},\qquad
\be(r)\sim 1/r.
\label{nine}
\eeq
The asymptotic form of $\be$ is a pure gauge, and describes the total magnetic flux, which can be seen to be proportional to the charge by integrating the second of Eqs. \eqref{six}.

There is a restriction on the frequency $\omega$ for which {\qb}s may exist, which can be seen by looking at the ungauged version of the first of Eqs. \eqref{six}. The $\omega$-dependent term can be thought of as being a part of an effective potential, and (as explained in \cite{Coleman:1985ki}) for {\qb}s to exist,
\beq
\sqrt{1-{3\over16g}}<\om<1.
\label{ten}
\eeq
For ungauged {\qb}s, values of $\om$ near the upper end of this range correspond to small {\qb}s, while the lower end of the range corresponds to large {\qb}s. Things are slightly more complicated in the gauged case (whether with or without a Chern-Simons term), as will be described below.

The equations of motion have four parameters: $e,\ g,\ \ka,\ \om$. Of these, the first three are parameters of the model itself (appearing in \eqref{one}), while the fourth is a parameter of the ansatz. A potentially interesting limit of the model is the pure Chern-Simons case. This can be realized by the change of variables $\bar A_\mu\equiv e A_\mu$ and $\bar\kappa\equiv\kappa/e^2$, after which the only appearance of $e$ in the Lagrangian is in the first term, which becomes $-\bar F_{\mu\nu}^2/4e^2$. The Maxwell term is then eliminated by setting $e\to\infty$ with $\bar\kappa$ fixed. We have not studied this limit here, although our analysis suggests that {\qb}s would require a very large value of $\bar\kappa$.

Once a solution is found, the energy and charge are evaluated using \eqref{seven}, and the ratio of the two indicates whether decay to ordinary matter is energetically possible or not.

Of particular interest is the lowest value of $E/Q$ for given parameters of the model, since this indicates the maximal energy savings gained in forming a {\qb}; it is thus an indicator of the most stable configuration (at least, among {\qb}s). Let the charge which minimizes $E/Q$ be $Q_{\rm min}$. If $Q_{\rm min}$ is zero or infinity, then {\qb}s would tend to break apart or coalesce, respectively. For finite values of $Q_{\rm min}$, it is easy to see that several {\qb}s of charge less than $Q_{\rm min}$ would tend to coalesce to one or more {\qb}s of charge $Q_{\rm min}$, while a {\qb} of greater charge could reduce its energy by giving off {\qb}s of charge $Q_{\rm min}$. (A more precise statement would require more detailed knowledge of $E/Q$ as a function of $Q$.)

Given that the model has three parameters, a complete exploration of parameter space would be rather involved. Rather than do this, since we are most interested in the effect of the Chern-Simons term on the existence and properties of {\qb}s, we have restricted ourselves to a specific value of $g$ (namely, $g=0.5$), chosen arbitrarily, apart from the fact that it does fall into the allowed range given above.

What can be said qualitatively, before attacking the problem numerically? Even in a gauged model without Chern-Simons term \cite{Lee:1988ag}, a qualitative analysis is much more difficult than in the ungauged case because, with the addition of the gauge field, Coleman's mechanical analogy no longer applies. This is all the more so in the MCS case, where the ansatz has three fields. However, a couple of general observations can easily be made. 

Suppose we held $Q$ fixed and ``turned on" $e$. Clearly, this would create an additional contribution to the {\qb} energy coming from the electromagnetic field. Thus, we would expect $E/Q$ to increase with $e$ for fixed $Q$; furthermore, we would expect the maximal charge at which {\qb}s occur to decrease with increasing $e$, very similar to the gauged case \cite{Lee:1988ag}.

As for the Chern-Simons term, as mentioned above, its presence is essential (in two dimensions) since without it the {\qb}'s energy diverges, so we cannot turn its coefficient $\ka$ on, since it cannot be zero. How do we expect the energetics to vary as the coefficient of the Chern-Simons term varies? Since the gauge field's mass 
is proportional to $\ka$, the electromagnetic fields have a decay length of $1/\ka$, so as $\ka$ increases, the electromagnetic contribution to the {\qb} energy should decrease, and {\qb}s should be stable over a wider range of the other parameters (or, equivalently, over a wider range of charges). 

These two qualitative behaviours are indeed seen in our numerical work, as will now be discussed.

The search for {\qb}s was performed using an iterative method known as relaxation, as described in detail in \cite{Press:1992}. A discretized configuration is provided as an initial guess; the algorithm estimates an error by determining to what extent this configuration is {\em not} a solution, adds a correction to the configuration in an intelligently-chosen ``direction" in configuration space, re-estimates the error, and so on, until the error is sufficiently small.


Note that the relaxation process is in no sense an actual physical evolution of the system. In particular, the charge is not conserved from one iteration to the next. The method attempts to find an approximate solution, given the four parameters $e,\ g,\ \ka,\ \om$. A more physical approach would be to specify $e,\ g,\ \ka$ and the charge $Q$ and calculate the minimum-energy solution for fixed $Q$; however, $Q$ and $\om$ are inextricably linked, and we could not replace one by the other.

\begin{figure}[ht]
\begin{center}
\begin{tabular}{cc}
\includegraphics[width=7cm]{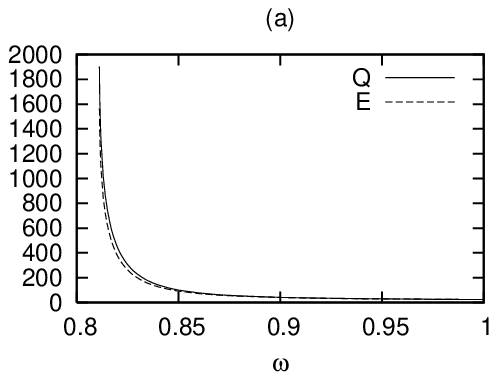} &
\includegraphics[width=7cm]{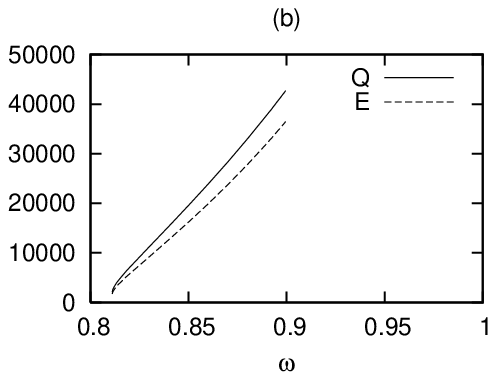} \\
\includegraphics[width=7cm]{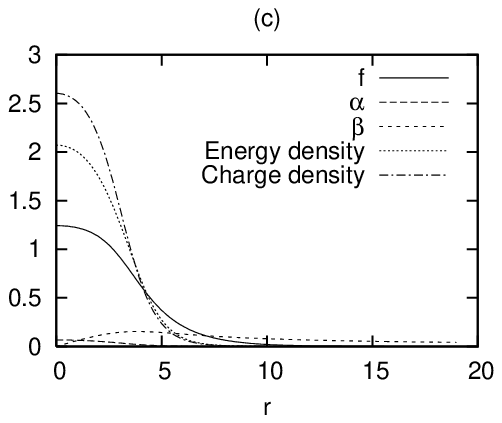} &
\includegraphics[width=7cm]{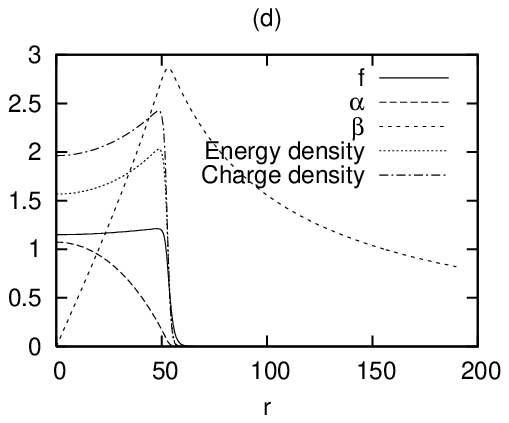}
\end{tabular}
\end{center}
\caption{$Q$ and $E$ vs $\om$ for (a) small and (b) large {\qb}s, at parameter values $(e,g,\ka)=(0.1,0.5,2.0)$. Typical (c) small and (d) large {\qb} profiles, at $\om=0.85$. Their charges are 99.998 and 19557, respectively.}
\label{figone}
\end{figure}

Let us describe in detail our results for typical values of the parameters of the model: $(e,g,\ka)=(0.1,0.5,2.0)$. Varying $\om$ within the allowed range given in \eqref{ten} (for $g=0.5$, this is $0.7906\le\om\le1$) reveals two classes of \qb, which can be described as small and large {\qb}s. Small {\qb}s, with charges ranging from about 20 to roughly 2000, exist in the range $0.8111\leq\om<1.0$, while large {\qb}s, with charges ranging from about 2000 to 43000, exist for $0.8111\leq\om\leq0.8995$ (Fig.~\ref{figone}). For $\om$ in the tiny region $0.7906\le\om\le0.8111$, no {\qb} solutions were found.

Also displayed in Fig.~\ref{figone} are typical small and large {\qb}s. Note that, as was the case with gauged {\qb}s \cite{Lee:1988ag}, the charge density of large {\qb}s increases from the centre to the exterior before dropping off, a behaviour which can be attributed to the repulsive gauge force. Note also that the prominent ``tail" of the field $\beta$ in Fig.~\ref{figone}d is a pure gauge, as mentioned above; all physical quantities tend to zero exponentially as $r\to\infty$, as expected.

The distinction between {\em small} and {\em large} {\qb}s is not arbitrary. Indeed, note the substantial overlap of the regions where small and large {\qb}s are found: for $0.8111\leq\om\leq0.8995$, this was the case.
For any $\om$ in this overlap region, {\qb}s of two different charges exist. For instance, at $\om=0.85$, the charges turn out to be very close to 100 and 20000. As $\om$ decreases, the difference between the charges decreases, tending toward zero as $\om$ approaches 0.8111, though numerical instability left a tiny gap (which we believe to be a numerical artifact) between the two.

\begin{figure}[ht]%
\begin{center}
\begin{tabular}{cc}
\includegraphics[width=7cm]{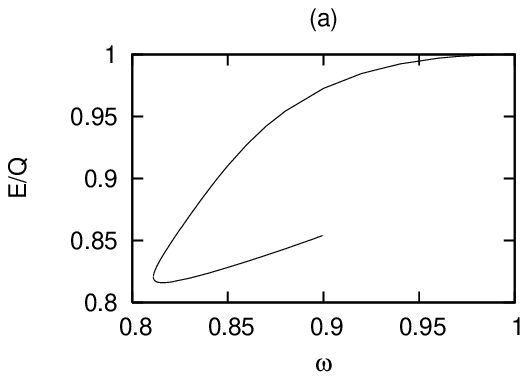} &
\includegraphics[width=7cm]{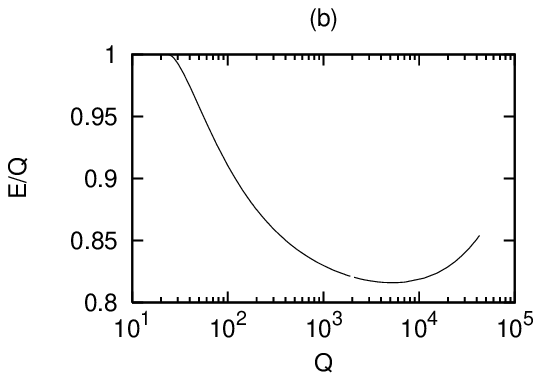}
\end{tabular}
\end{center}
\caption{$E/Q$ vs (a) $\om$ and (b) $Q$, at parameter values $(e,g,\ka)=(0.1,0.5,2.0)$. Note the tiny gap between the upper and lower branches of (a) (corresponding to small  and large {\qb}s, respectively), and the left and right branches of (b) (idem).}
\label{figtwo}
\end{figure}

Since $E/Q\ge1$ for ordinary matter, the energetic advantage of forming a {\qb} can be seen by comparing $E$ and $Q$. From Figs.~\ref{figone}a,b, it is apparent that large {\qb}s are more advantageous than small {\qb}s; this can be seen more directly by plotting $E/Q$ as a function of $\om$ and of $Q$ (Fig.~\ref{figtwo}). The tiny gap in these figures is the numerical artifact just mentioned, due to the numerically delicate transition between small and large {\qb}s.

Nothing particularly dramatic occurs in the extreme cases; for instance, as $\om$ approaches unity (the maximum allowed value), the {\qb} profile tends smoothly towards a nontrivial but qualitatively unremarkable limit; its charge approaches the smallest {\qb} charge found, namely, about 23.63. In this same limit, the energetic advantage, $E/Q$, approaches 1 from below, so these {\qb}s are only marginally preferable over ordinary matter. Similarly, at the large end of the large {\qb} curve, no dramatic change occurred to the solution found; from one value of $\om$ to the next the solution simply disappeared. No amount of coaxing (for instance, changing $\om$ extremely slowly and using the previous solution as initial guess) could entice the program to converge. We also attempted to study these endpoints using a different method (shooting, also described in \cite{Press:1992}), to no avail.

Two features of our results are unexpected. The first concerns the fact that, while $Q$ is a monotonic decreasing function of $\om$ in the ungauged case, this is not the case of MCS {\qb}s: small {\qb}s act in this way, but the charge of large {\qb}s is a monotonic {\em increasing} function of $\om$. Indeed, the very existence of two {\qb}s at the same value of $\om$ is unlike the ungauged case.

The second unexpected feature of our results concerns the maximum {\qb} charge. As explained above, while there is no upper limit to the ungauged {\qb}'s charge, we anticipate a maximal charge in the gauged case (with or without the Chern-Simons term), due to the electromagnetic contribution to the {\qb}'s energy. This contribution, one would think, should give rise to $E/Q>1$, at which point {\qb}s are no longer energetically advantageous compared with ordinary matter. Indeed there {\em is} a maximal charge. However, as can be seen from Fig.~\ref{figtwo}, this occurs when $E/Q$ is increasing but nonetheless considerably less than one. (At the maximal charge, $E/Q=0.8540$). This is also the case with gauged {\qb}s in three dimensions, as can be seen from Fig.~3 of \cite{Lee:1988ag}. Perhaps there is another explanation for this maximal charge (other than the inability to compete energetically with ordinary matter), but we have not come up with one. 

\begin{figure}[ht]%
\begin{center}
\begin{tabular}{cc}
\includegraphics[width=7cm]{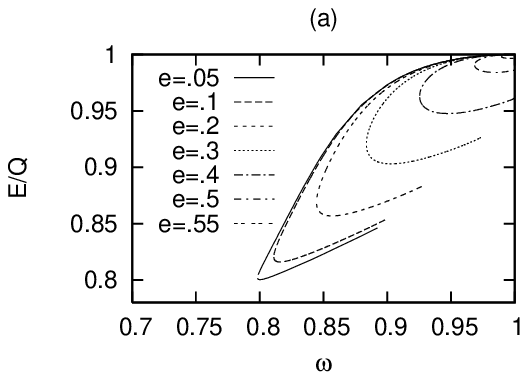} &
\includegraphics[width=7cm]{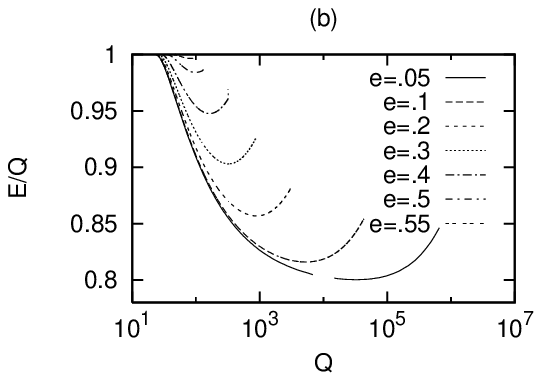}
\end{tabular}
\end{center}
\caption{$E/Q$ vs (a) $\om$ and (b) $Q$, for various values of $e$, with $(g,\ka)=(0.5,2.0)$.}
\label{figthree}
\end{figure}

Let us now discuss the effect of varying $e$ and $\ka$ on {\qb} energetics, beginning with $e$. As mentioned above, increasing $e$ is expected to give rise to a greater electromagnetic contribution to the {\qb} energy, reducing the range over which they exist.  This behaviour is borne out by our analysis, as illustrated in Fig.~\ref{figthree}, where $E/Q$ is plotted as a function of $\om$ and of $Q$ for several values of $e$. At the largest value displayed ($e=0.5$), {\qb}s exist only for $\om>0.9685$, and the maximal charge is only about 130. (To compare, the maximal charge at $e=0.05$ is about 650,000!)

\begin{figure}[ht]%
\begin{center}
\begin{tabular}{cc}
\includegraphics[width=7cm]{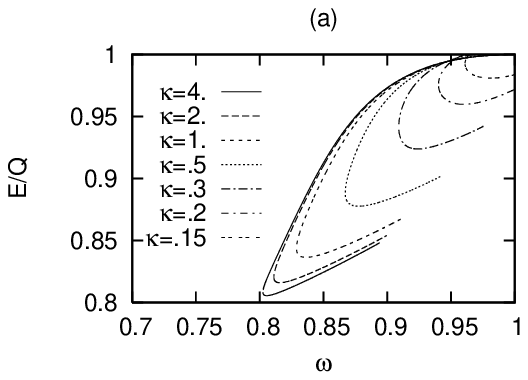} &
\includegraphics[width=7cm]{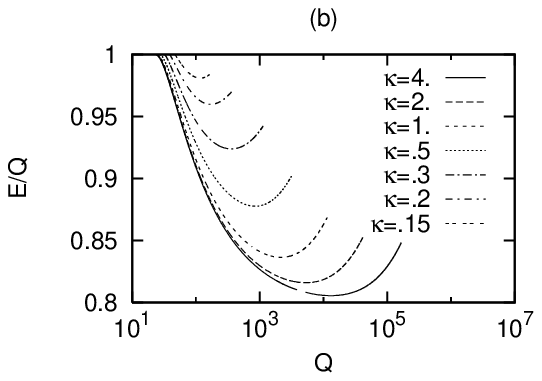}
\end{tabular}
\end{center}
\caption{$E/Q$ vs  (a) $\om$ and (b) $Q$, for various values of $\ka$, with $(e,g)=(0.1,0.5)$.}
\label{figfour}
\end{figure}

As for varying $\ka$, as we have already argued above, we expect that increasing $\ka$ (with all else fixed) will reduce the electromagnetic contribution to the {\qb} energy, and will therefore lead to an increase in the range and stability of {\qb}s. Again, our analysis confirms this expectation, as shown in Fig.~\ref{figfour}.

We are now in a position to speculate on what is to be expected in the pure Chern-Simons limit. This corresponds to $e\to\infty$ with $\kappa/e^2$ fixed. Figures \ref{figthree} and \ref{figfour} indicate that {\qb}s offer the greatest energy advantage for small $e$ and large $\kappa$; these suggest that $\bar\kappa$ must be quite large in order for {\qb}s to be energetically advantageous in this limit. This speculation could be tested with further work.

To summarize, we have argued that in two space dimensions, a Chern-Simons term (or some other mass generation mechanism for the gauge field) is necessary in order for {\qb}s to have finite energy. Considering the simplest model which might then give rise to {\qb}s, we have performed a numerical search for these objects. Two types were found for a wide range of parameters, large and small {\qb}s, the former being more stable in that they typically have lower values of $E/Q$. While many aspects of the behaviour of these objects are largely as expected, the fact that large {\qb}s cease to exist when they do (in particular, with $E/Q$ well below unity) is surprising and merits further study.

\bigskip
We thank Manu Paranjape for interesting conversations. This work was
funded in part by the
National Science and Engineering Research Council.


\bibliographystyle{unsrt}
\bibliography{fieldtheory}

\begin{thebibliography}{10}

\bibitem{Lee:1991ax}
T.~D. Lee and Y.~Pang.
\newblock Nontopological solitons.
\newblock {\em Phys. Rept.}, 221:251--350, 1992.

\bibitem{Coleman:1985ki}
Sidney~R. Coleman.
\newblock Q balls.
\newblock {\em Nucl. Phys.}, B262:263, 1985.

\bibitem{Kusenko:1997ad}
Alexander Kusenko.
\newblock Small {Q} balls.
\newblock {\em Phys. Lett.}, B404:285, 1997.

\bibitem{Kusenko:1999gz}
Alexander Kusenko.
\newblock Cosmology of {S}usy {Q}-balls.
\newblock 1999.

\bibitem{Hong:1987ur}
Deog~Ki Hong.
\newblock Q balls in superfluid {He}-3.
\newblock {\em J. Low. Temp. Phys.}, 71:483, 1988.

\bibitem{Enqvist:2003zb}
K.~Enqvist and M.~Laine.
\newblock Q-balls in atomic {Bose-Einstein} condensates.
\newblock {\em JCAP}, 0308:003, 2003.

\bibitem{Nussinov:2004hf}
Zohar Nussinov and Shmuel Nussinov.
\newblock Non-relativistic {Bose-Einstein} condensates, kaon droplets, and
  {Q}-balls.
\newblock 2004.

\bibitem{Safian:1987pr}
Alexander~M. Safian, Sidney~R. Coleman, and Minos Axenides.
\newblock Some nonabelian {Q} balls.
\newblock {\em Nucl. Phys.}, B297:498, 1988.

\bibitem{Safian:1988cz}
Alexander~M. Safian.
\newblock Some more nonabelian {Q} balls.
\newblock {\em Nucl. Phys.}, B304:392--402, 1988.

\bibitem{Lee:1988ag}
Ki-Myeong Lee, Jaime~A. Stein-Schabes, Richard Watkins, and Lawrence~M. Widrow.
\newblock Gauged {Q} balls.
\newblock {\em Phys. Rev.}, D39:1665, 1989.

\bibitem{Anagnostopoulos:2001dh}
K.~N. Anagnostopoulos, M.~Axenides, E.~G. Floratos, and N.~Tetradis.
\newblock Large gauged {Q} balls.
\newblock {\em Phys. Rev.}, D64:125006, 2001.

\bibitem{Lynn:1988rb}
B.~W. Lynn.
\newblock Q stars.
\newblock {\em Nucl. Phys.}, B321:465, 1989.

\bibitem{Prikas:2004fx}
Athanasios Prikas.
\newblock Q-stars in extra dimensions.
\newblock {\em Phys. Rev.}, D69:125008, 2004.

\bibitem{Gleiser:2005iq}
Marcelo Gleiser and Joel Thorarinson.
\newblock Energy landscape of d-dimensional {Q}-balls.
\newblock 2005.

\bibitem{MacKenzie:2001av}
R.~B. MacKenzie and Manu~B. Paranjape.
\newblock From {Q}-walls to {Q}-balls.
\newblock {\em JHEP}, 08:003, 2001.

\bibitem{Axenides:2001pi}
M.~Axenides, E.~Floratos, S.~Komineas, and L.~Perivolaropoulos.
\newblock Q rings.
\newblock {\em Phys. Rev. Lett.}, 86:4459--4462, 2001.

\bibitem{Volkov:2002aj}
Mikhail~S. Volkov and Erik Wohnert.
\newblock Spinning {Q}-balls.
\newblock {\em Phys. Rev.}, D66:085003, 2002.

\bibitem{Kleihaus:2005me}
Burkhard Kleihaus, Jutta Kunz, and Meike List.
\newblock Rotating boson stars and {Q}-balls.
\newblock {\em Phys. Rev.}, D72:064002, 2005.

\bibitem{Prange:1990}
Richard~E. Prange and Steven~M. Girvin.
\newblock {\em The Quantum Hall Effect, 2nd Edition}.
\newblock Springer-Verlag, 1990.

\bibitem{Wilczek:1990ik}
Frank Wilczek.
\newblock {\em Fractional statistics and anyon superconductivity}.
\newblock World Scientific, 1990.

\bibitem{MacKenzie:1988ft}
R.~MacKenzie and Frank Wilczek.
\newblock Peculiar spin and statistics in two space dimensions.
\newblock {\em Int. J. Mod. Phys.}, A3:2827, 1988.

\bibitem{Goldhaber:1988iw}
Alfred~S. Goldhaber, R.~MacKenzie, and Frank Wilczek.
\newblock Field corrections to induced statistics.
\newblock {\em Mod. Phys. Lett.}, A4:21, 1989.

\bibitem{Redlich:1983kn}
A.~N. Redlich.
\newblock Gauge noninvariance and parity nonconservation of three- dimensional
  fermions.
\newblock {\em Phys. Rev. Lett.}, 52:18, 1984.

\bibitem{Wilczek:1983cy}
Frank Wilczek and A.~Zee.
\newblock Linking numbers, spin, and statistics of solitons.
\newblock {\em Phys. Rev. Lett.}, 51:2250, 1983.

\bibitem{Arovas:1985yb}
D.~P. Arovas, J.~R. Schrieffer, Frank Wilczek, and A.~Zee.
\newblock Statistical mechanics of anyons.
\newblock {\em Nucl. Phys.}, B251:117--126, 1985.

\bibitem{Deser:1981wh}
S.~Deser, R.~Jackiw, and S.~Templeton.
\newblock Topologically massive gauge theories.
\newblock {\em Ann. Phys.}, 140:372--411, 1982.

\bibitem{Khare:1989dx}
Avinash Khare and Sumathi Rao.
\newblock Q ball solutions in local gauge theories with {C}hern-{S}imons term.
\newblock {\em Phys. Lett.}, B227:424, 1989.

\bibitem{Khare:1990jz}
Avinash Khare.
\newblock Charged vortices and {Q} balls in an abelian {H}iggs model exhibiting
  first order phase transition.
\newblock {\em Phys. Lett.}, B255:393--397, 1991.

\bibitem{Jackiw:1990pr}
R.~Jackiw, Ki-Myeong Lee, and Erick~J. Weinberg.
\newblock Selfdual chern-simons solitons.
\newblock {\em Phys. Rev.}, D42:3488--3499, 1990.

\bibitem{Lee:1990eq}
Choon-kyu Lee, Ki-Myeong Lee, and Hyunsoo Min.
\newblock Selfdual maxwell chern-simons solitons.
\newblock {\em Phys. Lett.}, B252:79--83, 1990.

\bibitem{Jackiw:1991au}
R.~Jackiw and So-Young Pi.
\newblock Selfdual chern-simons solitons.
\newblock {\em Prog. Theor. Phys. Suppl.}, 107:1--40, 1992.

\bibitem{Press:1992}
William~H. Press, Saul~A. Teukolsky, William~T. Vetterling, and Brian~P.
  Flannery.
\newblock {\em Numerical Recipes in Fortran 77: The Art of Scientific
  Computing, Second Edition}.
\newblock Cambridge University Press, 1992.

\end{thebibliography}


\end{document}